\documentclass[12pt,a4paper]{article} 
\usepackage{latexsym}
\usepackage{epsfig}
\renewcommand{\thefootnote}{\#\arabic{footnote}}
\begin{document}
{\vspace*{-3cm} \normalsize \hfill
    \parbox{38mm}{MS-TPI-00-11\\
                  hep-lat/0012005}}
\vspace*{30mm}
\begin{center}
\begin{Large} \begin{bf}
Lattice Gauge Theory \\
A Short Primer\\
\end{bf}  \end{Large}
\vspace*{1cm}
\begin{large}
Lectures given at the PSI Zuoz Summer School 2000\\
\end{large}
\vspace*{2cm}
\begin{large}
\renewcommand{\thefootnote}{\fnsymbol{footnote}}
\setcounter{footnote}{0}
G.~M\"unster\,\footnote[1]{{\it Electronic mail address:}
{\tt munsteg@uni-muenster.de}},\
M.~Walzl\,\footnote[2]{{\it Electronic mail address:}
{\tt m.walzl@fz-juelich.de}}\\
\setcounter{footnote}{0}
\end{large}
\medskip
$\ast$Institut f\"ur Theoretische Physik\\
Westf\"alische Wilhelms-Universit\"at\\
Wilhelm-Klemm-Str. 9\\
D-48149 M\"unster\\
Germany\\
\medskip
$\dagger$Institut f\"ur Kernphysik (Theorie)\\
Forschungszentrum J\"ulich\\
D-52425 J\"ulich\\
Germany
\end{center}
\vspace*{5mm}
\begin{center}
{\bf Abstract}
\end{center}
\begin{quotation}
\noindent
In this contribution we give an introduction to the foundations and
methods of lattice gauge theory.  Starting with a brief discussion of
the quantum mechanical path integral, we develop the main ingredients
of lattice field theory:  functional integrals, Euclidean field theory
and the space-time discretization of scalar, fermion and gauge fields.
Some of the methods used in calculations are reviewed and illustrated by
a collection of typical results.
\end{quotation}
\section{Introduction}

\subsection{Why this Article?}

These lectures given at the PSI summer school 2000 in Zuoz give an overview
of the basic ideas and results of lattice gauge theory for non-experts, with
a stress on lattice quantum chromodynamics.  This is neither a review of the
status nor a survey of recent results in lattice gauge theory.  Some typical
results are presented for illustrative purposes only.  For reviews on recent
developments see \cite{reviews} or the proceedings of the annual LATTICE
conferences.  To those who ask for a more detailed and in some places more
rigorous look at this subject we recommend the literature in
\cite{literature}.

\pagestyle{plain}

The article is organized as follows.  We start with a motivation of
non-perturbative treatments of gauge theories (especially QCD).  The second
chapter is an introduction to Euclidean field theory with bosons. We discuss
functional integrals with imaginary time coordinates and space time
discretization on a lattice at some length.  Next, we sketch how to
implement gauge fields on a space-time lattice, before we discuss methods
used to evaluate functional integrals.  In order to perform calculations in
Quantum Chromodynamics (QCD), a lattice treatment of fermions is necessary,
and in the fifth chapter we describe ways to put fermions on a lattice.  We
close with an overview of physical problems that can be addressed.  Some
interesting results are shown, the problem of how to approach the continuum
limit is discussed and error sources are listed.

\subsection{Why the Lattice?}

To answer this question it is instructive to focus on a prominent example
for gauge theories on a space-time lattice: QCD. Therefore let us briefly
highlight the major properties of this theory: QCD originated historically
as a development of the quark model.  In 1961 Gell-Mann and Ne'eman
established a classification scheme for the hadrons known in those days,
using representations of what we call SU(3)$_{\mbox{\scriptsize flavour}}$
today.  This model has one remarkable feature: the fundamental
representations (typical notation is $3$ and $\bar3$) did not appear to
be realized in nature.  Besides other hints this led Gell-Mann and Zweig to
introduce hadrons as composite objects, with particles in the fundamental
representations (the quarks and antiquarks) as their constituents.  Now the
quark model allows two different combination of quarks and antiquarks to
exist:

\begin{equation}
\mbox{Baryons:}\qquad 3 \otimes 3 \otimes 3 = 10 \oplus 8 \oplus
8 \oplus 1
\end{equation}
\begin{equation}
\mbox{Mesons:}\qquad 3 \otimes \bar3 = 8 \oplus 1.
\end{equation}

Instead of going into the details of this kinematical quark model we ask for
the dynamics that allows these particular combinations: why are quarks
confined in baryons and mesons, what is the mechanism that forbids the
particles in the fundamental representations to be free?

\begin{figure}[hbt]
\begin{center}
\epsfig{file=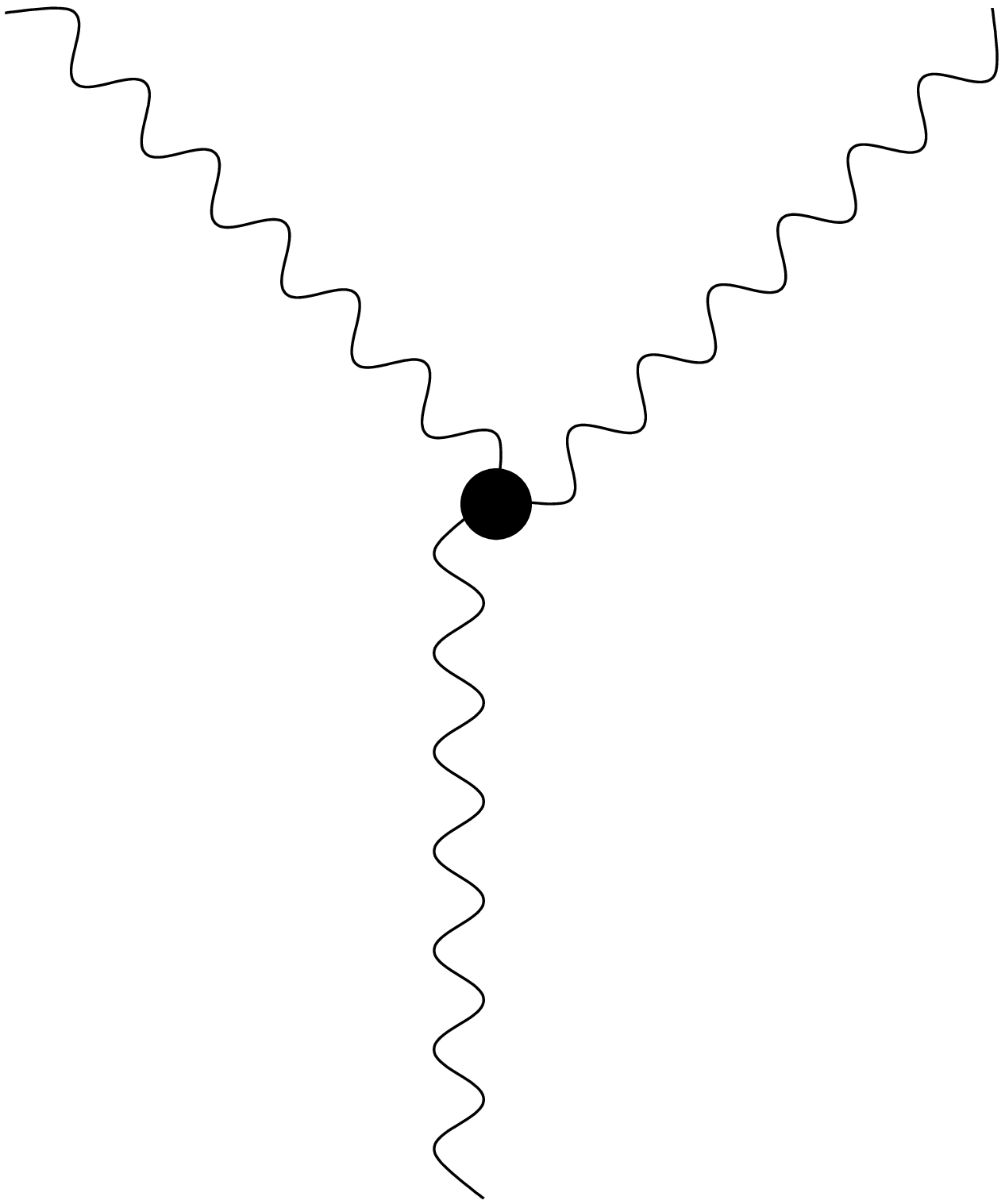,height=2cm}
\hspace{8mm}
\epsfig{file=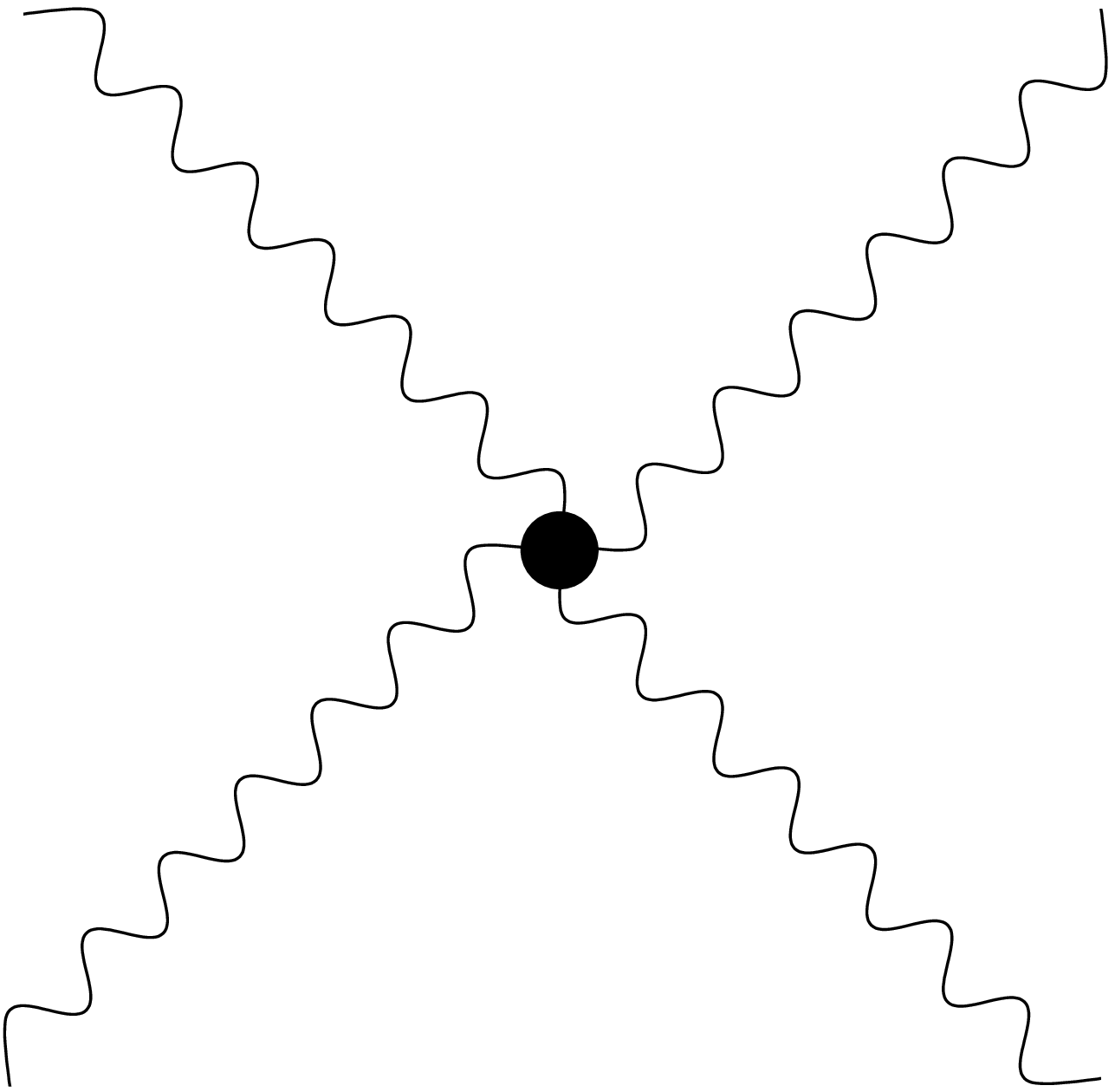,height=2cm}
\caption{Gluon self--interactions}
\label{selfinteractions}
\end{center}
\end{figure}

\begin{figure}[hbt]
\begin{center}
\epsfig{file=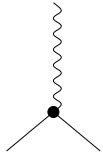,height=2cm}
\caption{Quark--gluon interaction}
\label{photonscatter}
\end{center}
\end{figure}

The Standard Model contains QCD as a theory for quark dynamics. This gauge
theory with SU(3)$_{\mbox{\scriptsize colour}}$ as a gauge group introduces
colour as a degree-of-freedom for the quarks (again, the quarks appear in
the fundamental representation) and eight mediating vector bosons (the
gluons) in the adjoint representation (they always carry two colour
degrees-of-freedom and can be expanded in a basis of the eight Gell-Mann
matrices).
\begin{equation}
\mbox{quark fields:}\hspace{2mm} q_{i}^{f}(x),\quad
i=1,2,3,\ f=1,\ldots,N_f
\end{equation}
\begin{equation}
\mbox{gluon fields:}\hspace{2mm} A_{\mu}^{a}(x),\quad a=1,\ldots,8.
\end{equation}
Here $i$ and $a$ are colour indices according to the relevant representation
of SU(3)$_{\mbox{\scriptsize colour}}$, while $f$ labels the quark flavours
u,d,s,...,$N_f$, and $\mu$ is the Lorentz index.  The analogy to Quantum
Electrodynamics (QED) (colour $\leftrightarrow$ charge and gluons
$\leftrightarrow$ photons) is violated by the existence of three different
colours in QCD in contrast to only one charge in QED, which reflects the
different nature of the gauge groups SU(3)$_{\mbox{\scriptsize colour}}$
and U(1)$_{\mbox{\scriptsize hypercharge}}$.  While
U(1)$_{\mbox{\scriptsize hypercharge}}$ is Abelian,
SU(3)$_{\mbox{\scriptsize colour}}$ is not. As a consequence QCD
exhibits self-interaction vertices, see Fig.~\ref{selfinteractions}, in
addition to the QED-like quark-gluon interactions shown in
Fig.~\ref{photonscatter}.


An important property of QCD is {\em asymptotic freedom}.  The interaction
strength depends on the typical energy scale $Q^2$ involved in the
interaction process under consideration.  The running of the coupling
constant can be determined using renormalization group equations, with the
result:
\begin{equation}
g_{QCD}^2(Q^2) = \frac{1}{\beta_0 \log(Q^2 / \Lambda^2)} + \cdots
\end{equation}
with $\Lambda \approx 1$ GeV and
\begin{equation}
\beta_0 = \frac{33 - 2 N_f}{48 \pi^2} \ge 0,
\end{equation}
in contrast to QED, where $\beta_0 \le 0$.

The coupling decreases with energy in QCD.  As a consequence, perturbation
theory, which amounts to expansions in powers of the coupling, is
phenomenologically successful for large energies, i.e.\
$Q^2 \gg \Lambda^2$.  The predictive power of QCD at high energies will
be treated in Stirling's lectures \cite{Stirling}.

The question why quarks are confined is a question to the low energy regime
of QCD and appears to require techniques beyond perturbation theory.
There are other interesting low energy questions:
\begin{itemize}
\item {\em prediction of the hadron spectrum}: dynamical mass generation
\item {\em hadron properties}: wave-functions, matrix-elements, \dots
\item {\em hadronization} in  deep inelastic scattering\
(jet formation, \dots)
\item {\em chiral symmetry breaking}\ (expectation values for quark
condensates, low energy constants of effective Lagrangians, \dots)
\item {\em energy scale dependence of the coupling at low $Q^2$}.
\end{itemize}
These are examples of non-perturbative problems in QCD.

Not only QCD, but also other components of the Standard Model and moreover
theories of physics beyond the Standard Model supply us with
non-perturbative problems:
\begin{itemize}
\item {\em Higgs-Yukawa models}: parameters of the Standard Model\
(predictions for the Higgs mass, CKM - Matrix, \dots)
\item {\em QED}\ (new phases, \dots)
\item {\em quantum gravity and SUSY}.
\end{itemize}
An important step to answer such questions has been made by K.~Wilson in
1974 \cite{Wilson}.  He introduced a formulation of QCD on a space-time
lattice, which allows the application of various non-perturbative
techniques.  In the following chapters we shall explain this discretization
in detail.  It leads to mathematically well-defined problems, which are (at
least in principle) solvable.  What can be achieved in practice will be
discussed below.  It should also be pointed out that the introduction of a
space-time lattice can be taken as a starting point for a mathematically
clean approach to quantum field theory, so-called constructive quantum field
theory.

\section{Quantum Fields on a Lattice}

This chapter is an introduction to the main concepts of lattice gauge
theory:  quantum field theory in its path integral formulation, Wick
rotation to imaginary time coordinates leading to Euclidean field theory,
and the discretization of space-time in form of a lattice.
We shall illustrate these concepts with a scalar field theory.

\subsection{The Quantum Mechanical Path Integral}

\begin{figure}[hbt]
\begin{center}
\epsfig{file=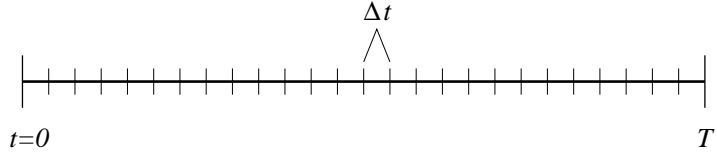,height=2cm}
\caption{Discretized time interval}
\label{Tpart}
\end{center}
\end{figure}

To illuminate this crucial construction let us first consider the case of
quantum mechanics of a particle in one space dimension.
Let the Hamiltonian be
\begin{equation}
H = \frac{p^2}{2m} + V(x) \equiv H_0 + V.
\end{equation}
The quantum mechanical transition amplitude is
\begin{equation}
\langle x',t'|x,t\rangle = \langle x'|e^{-iH(t'-t)}|x\rangle.
\end{equation}
Inserting a complete set of (improper) coordinate eigenstates,
\begin{equation}
1 = \int\!dx_1 \, |x_1\rangle\langle x_1|,
\end{equation}
into the matrix element, taking $T = (t'- t)$ and
$\Delta t = (t_1 - t)$,
we obtain
\begin{equation}
\langle x',t'|x,t\rangle = \int\!dx_1\,\langle x'|e^{-iH(T-\Delta t)}
|x_1\rangle \langle x_1|e^{-iH\Delta t}|x\rangle.
\end{equation}
Dividing $T$ into $n$ equal parts, $T=n\Delta t$, as shown in
Fig.~\ref{Tpart}, and inserting $(n-1)$ complete sets in this way, one
gets
\begin{eqnarray}
\lefteqn{\langle x',t'|x,t\rangle =}\nonumber\\
&&\hspace*{-7mm}
\int\!dx_1\dots dx_{n-1}\ \langle x'|e^{-iH\Delta t}|x_{n-1}\rangle
\langle x_{n-1}|e^{-iH\Delta t}|x_{n-2}\rangle\dots
\langle x_1|e^{-iH\Delta t}|x\rangle.
\end{eqnarray}
In the following we set $x \equiv x_0$ and $x' \equiv x_n$.

\begin{figure}[hbt]
\begin{center}
\epsfig{file=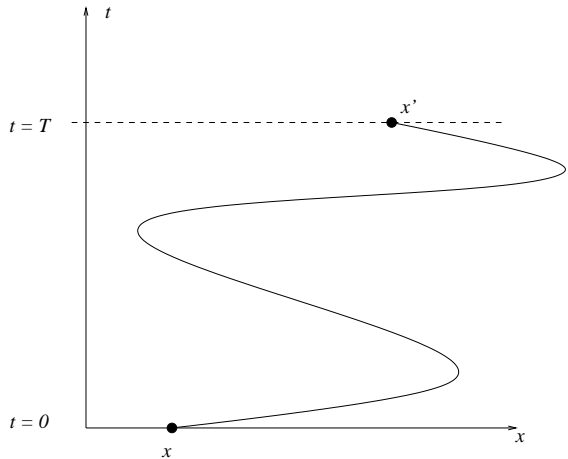,height=6cm}
\caption{Path of a particle}
\label{path}
\end{center}
\end{figure}

For large $n$, when $\Delta t$ becomes small, we can rewrite the matrix
elements by using only the first term of the
Baker-Campbell-Hausdorff-formula as a good approximation to the exponential:
\begin{eqnarray}
\lefteqn{\langle x_{k+1}|e^{-iH\Delta t}|x_{k}\rangle \approx} \nonumber\\
&\,&\langle x_{k+1}|e^{-iH_0\Delta t}e^{-iV\Delta t}|x_{k}\rangle\,
= \langle x_{k+1}|e^{-iH_0\Delta t}|x_{k}\rangle\,e^{-iV(x_k)\Delta t},
\end{eqnarray}
using the fact that $V$ only depends on the space coordinates.
The remaining matrix element can be calculated by means of Fourier transform
with the result
\begin{equation}
\langle x_{k+1}|e^{-iH\Delta t}|x_{k}\rangle \approx
\sqrt{\frac{m}{2\pi i\Delta t}} \ \exp i\Delta t \left\{ \frac{m}{2}
\left( \frac{x_{k+1} - x_k}{\Delta t} \right)^2 - V(x_k) \right\}.
\end{equation}
Doing so for every matrix element, the amplitude turns into
\begin{equation}
\langle x'|e^{-iHT}|x\rangle  = \int \frac{dx_1\dots dx_{n-1}}
{(\frac{2\pi i\Delta t}{m})^{n/2}} \exp\,i\sum^{n-1}_{k=0}\Delta t
\left\{ \frac{m}{2}\left(\frac{x_{k+1}-x_k}{\Delta t}\right)^2
- V(x_k)\right\}.
\end{equation}
What is this good for? In the limit $n\rightarrow \infty$ we observe that
the exponent becomes the classical action
\begin{eqnarray}
&\,&\sum^{n-1}_{k=0}\Delta t \left\{ \frac{m}{2}
\left(\frac{x_{k+1}-x_k}{\Delta t}\right)^2 - V(x_k) \right\} \nonumber\\
&\,& \longrightarrow \, \int_0^T\!dt \left\{
\frac{m}{2}\left(\frac{dx}{dt}\right)^2 - V(x)\right\}
= \int_0^T\!dt\,L(x,\dot x) \equiv S
\end{eqnarray}
for a path $x(t)$ from $x$ to $x'$ with $x_k = x(k\Delta t)$,
see Fig.~\ref{path}.

Second, notice that the integrations over the $x_k$ can be interpreted as
an integration of the system over all possible paths $x(t)$.
Therefore we introduce the notation
\begin{equation}
\left(\frac{m}{2\pi i\Delta t}\right)^{n/2}\ dx_1\dots dx_{n-1}
\longrightarrow \mbox{const.} \prod_t dx(t) \equiv {\cal D}x
\end{equation}
and arrive at the path integral representation of the quantum mechanical
amplitude:
\begin{equation}
\langle x'|e^{-iHT}|x\rangle = \int\!{\cal D}x \ e^{iS}.
\end{equation}
For a particle in 3 dimensional space we generalize to paths $x_i(t)$, where
$i$=1,2,3, and
\begin{equation}
{\cal D}x = \prod_t \prod_i dx_i(t) .
\end{equation}

Perhaps this is the most intuitive picture of the quantum mechanical
transition amplitude. It can be written as an integral over contributions
from all possible paths from the starting point to the final point. Each
path is weighted by the classical action evaluated along this path.

\subsection{Quantum Field Theory with Functional Integrals}

Now we are going to translate to field theory this representation of quantum
mechanics in terms of path integrals. We consider a scalar field $\phi(x)$,
where $x = (\vec x, t)$ labels space-time coordinates, and the time
evolution of $\phi(\vec x,t)$ is given by
\begin{equation}
\phi (\vec x,t) = e^{iHt}\,\phi (\vec x, t=0)\,e^{-iHt}.
\end{equation}
The objects of interest in field theory are vacuum expectation values of
(time ordered) products of field operators, the Greens functions:
\begin{equation}
\langle 0|\phi(x_1)\phi(x_2)\dots\phi(x_n)|0\rangle ,
\qquad t_1 > t_2 > \dots > t_n.
\end{equation}
Prominent examples are propagators
\begin{equation}
\langle 0 |\phi(x)\phi(y)|0\rangle.
\end{equation}
The Greens functions essentially contain all physical information.
In particular, S-matrix elements are related to Greens functions, e.g.\
the 2-particle scattering elements can be obtained from
\begin{equation}
\langle 0 | \phi(x_1) \dots \phi(x_4) | 0 \rangle.
\end{equation}

Instead of discussing the functional integral representation for quantum
field theory from the beginning, we shall restrict ourselves to translating
the quantum mechanical concepts to field theory by means of analogy. To this
end we would like to translate the basic variables $x_i(t)$ into fields
$\phi(\vec x,t)$. The rules for the translation are then
\begin{eqnarray*}
x_i(t) & \longleftrightarrow & \phi(\vec x,t)\\
i & \longleftrightarrow & \vec x\\
\prod_{t,i} dx_i(t) & \longleftrightarrow &
\prod_{t,\vec x} d\phi (\vec x,t) \equiv {\cal D}\phi\\
S = \int\!dt\ L & \longleftrightarrow & S = \int\!dt\,d^3x\ {\cal L},
\end{eqnarray*}
where $S$ is the classical action.

For scalar field theory we might consider the following Lagrangian
density:
\begin{eqnarray}
{\cal L} &=& \frac{1}{2}\left( (\dot\phi(x) )^2 - (\nabla \phi(x))^2\right)
- \frac{m_0^2}{2}\phi(x)^2 - \frac{g_0}{4!}\phi(x)^4\nonumber\\
&=& \frac{1}{2}(\partial_{\mu} \phi)(\partial^{\mu} \phi)
- \frac{m_0^2}{2}\phi(x)^2 - \frac{g_0}{4!}\phi(x)^4 .
\end{eqnarray}
The mass $m_0$ and coupling constant $g_0$ bear a subscript $0$, since they
are bare, unrenormalized parameters. This theory plays a role in the context
of Higgs-Yukawa models, where $\phi(x)$ is the Higgs field.

In analogy to the quantum mechanical path integral we now write down a
representation of the Greens functions in terms of what one calls
{\em functional integrals}:
\begin{equation}
\langle 0|\phi(x_1)\phi(x_2)\dots\phi(x_n)|0\rangle
= \frac{1}{Z} \int\!{\cal D}\phi\
\phi(x_1)\phi(x_2)\dots\phi(x_n)\,e^{iS}
\end{equation}
with
\begin{equation}
Z = \int\!{\cal D}\phi \ e^{iS}.
\end{equation}
These expressions involve integrals over all classical field configurations.

As mentioned before, we do not attempt any derivation of functional
integrals but just want to motivate their form by analogy. Furthermore, in
the case of quantum mechanics we considered the transition amplitude,
whereas now we have written the formula for Greens functions, which is a bit
different.

The formulae for functional integrals give rise to some questions. First of
all, how does the projection onto the groundstate $| 0 \rangle$ arise?
Secondly, these integrals contain oscillating integrands, due to the
imaginary exponents; what about their convergence? Moreover, is there a way
to evaluate them numerically?

In the following we shall discuss, how the introduction of imaginary times
helps in answering these questions.

\subsection{Euclidean Field Theory}

\begin{figure}[hbt]
\begin{center}
\epsfig{file=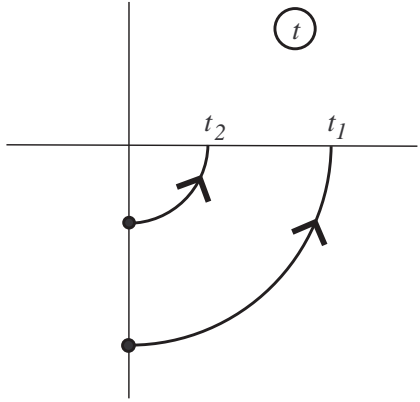,height=6cm}
\caption{Wick rotation from imaginary to real time coordinates}
\label{wickrot}
\end{center}
\end{figure}

Let us return to quantum mechanics for a moment. Here we can also introduce
Greens functions, e.g.
\begin{equation}
G(t_1,t_2) = \langle 0|X(t_1)X(t_2)|0\rangle,\qquad t_1 > t_2.
\end{equation}
We are now going to demonstrate that these Greens functions are related to
quantum mechanical amplitudes at imaginary times by analytic continuation.
Consider the matrix element
\begin{equation}
\langle x',t'|X(t_1)X(t_2)|x,t\rangle  = \langle x'|e^{-iH(t'-t_1)}
X e^{-iH(t_1-t_2)} X e^{-iH(t_2-t)}|x\rangle
\end{equation}
for $t' > t_1 > t_2 > t$.
Now let us choose all times to be purely imaginary
\begin{equation}
t = -i\tau,
\end{equation}
again ordered, $\tau' > \tau_1 > \tau_2 > \tau$. This yields the expression
\begin{equation}
\langle x'|e^{-H(\tau '-\tau_1)} X e^{-H(\tau_1-\tau_2)}
X e^{-H(\tau_2-\tau)}|x\rangle.
\end{equation}
Inserting a complete set of energy eigenstates, the expansion of the time
evolution operator in imaginary times is
\begin{equation}
e^{-H\tau} = \sum_{n=0}^{\infty} e^{-E_n\tau} |n\rangle
\langle n|
= |0\rangle \langle 0| + e^{-E_1\tau}|1\rangle \langle 1| + \dots,
\end{equation}
where the ground state energy has been normalized to $E_0 = 0$.
For large $\tau$ it reduces to the projector onto the groundstate.
Consequently, in the limit $\tau' \rightarrow \infty$ and
$\tau \rightarrow -\infty$ our matrix element becomes
\begin{equation}
\langle x'| 0 \rangle \langle 0 | X e^{-H(\tau_1-\tau_2)} X | 0 \rangle
\langle 0 | x \rangle,
\end{equation}
and similarly
\begin{equation}
\langle x'| e^{-H(\tau' - \tau)} | x \rangle \longrightarrow
\langle x' | 0 \rangle \langle 0 | x \rangle.
\end{equation}
Therefore the Greens function at imaginary times,
\begin{equation}
G_E(\tau_1,\tau_2) = \langle 0| Xe^{-H(\tau_1-\tau_2)} X |0\rangle,
\end{equation}
can be expressed as
\begin{equation}
G_E(\tau_1,\tau_2) =
\lim_{\tau'\rightarrow \infty \atop \tau \rightarrow -\infty}
\frac{\langle x'|e^{-H(\tau '-\tau_1)} X e^{-H(\tau_1-\tau_2)}
X e^{-H(\tau_2-\tau)}|x\rangle}{\langle x'|e^{-H(\tau'-\tau)}|x\rangle}.
\end{equation}
Now we can represent the denominator as well as the numerator by path
integrals as seen before. The difference is that for imaginary times we have
to use
\begin{equation}
\langle x | e^{-H\Delta \tau}| y \rangle \approx
\sqrt{\frac{m}{2\pi \Delta \tau}} \ \exp -\Delta \tau \left\{ \frac{m}{2}
\left( \frac{x - y}{\Delta \tau} \right)^2 + V(x) \right\}.
\end{equation}
This leads to the path integral representation
\begin{equation}
G_E(\tau_1,\tau_2) = \frac{1}{Z}\int\!{\cal D}x\ x(\tau_1) x(\tau_2) \, 
e^{-S_E},
\end{equation}
where
\begin{equation}
Z = \int\!{\cal D}x\ e^{-S_E}
\end{equation}
and
\begin{equation}
S_E = \int\!d\tau \left\{ \frac{m}{2} \left(\frac{dx}{d\tau}\right)^2 +
V(x(\tau))\right\}.
\end{equation}
The Greens function at real times, which we were interested in originally,
can be obtained from $G_E$ by means of analytical continuation, $G(t_1,t_2)
= G_E(i t_1, i t_2)$. The analytical continuation has to be done in such a
way that all time arguments are rotated simultaneously counter-clockwise
in the complex $t$-plane. This is the so-called Wick rotation,
illustrated in Fig.~\ref{wickrot}.

Now we turn to field theory again. The Greens functions
\begin{equation}
G(x_1,\dots,x_n) = \langle 0 | T \phi(x_1) \dots \phi(x_n)| 0 \rangle,
\end{equation}
continued to imaginary times, $t = -i\tau$, are the so-called Schwinger
functions
\begin{equation}
G_E( (\vec x_1,\tau_1), \dots, (\vec x_n,\tau_n)) =
G( (\vec x_1,-i\tau_1), \dots, (\vec x_n,-i\tau_n)).
\end{equation}
In analogy to the quantum mechanical case their functional integral
representation reads
\begin{equation}
G_E(x_1,\dots,x_n) = \frac{1}{Z}\int\!{\cal D}\phi\ \phi(x_1)\dots
\phi(x_n)\, e^{-S_E}
\end{equation}
with
\begin{equation}
Z = \int\!{\cal D}\phi\ e^{-S_E}
\end{equation}
and
\begin{eqnarray}
S_E &=& \int\!d^3xd\tau \left\{
\frac{1}{2}\left(\frac{d\phi}{d\tau}\right)^2
+ \frac{1}{2}(\nabla \phi)^2 + \frac{m_0^2}{2} \phi^2 + \frac{g_0}{4!}
\phi^4 \right\} \nonumber\\
&=& \int\!d^4x \left\{ \frac{1}{2}(\partial_{\mu}\phi)^2
+ \frac{m_0^2}{2}\phi^2 + \frac{g_0}{4!}\phi^4 \right\}.
\label{Eaction}
\end{eqnarray}
As can also be seen from the kinetic part contained in $S_E$, the metric
of Minkowski space
\begin{equation}
- ds^2 = -dt^2 + dx_1^2 + dx_2^2 + dx_3^2
\end{equation}
has changed into
\begin{equation}
d\tau^2 + dx_1^2 + dx_2^2 + dx_3^2 ,
\end{equation}
which is the metric of a Euclidean space. Therefore one speaks of {\em
Euclidean Greens functions} $G_E$ and of {\em Euclidean functional
integrals}. They are taken as starting point for non-perturbative
investigations of field theories and for constructive studies.

As $S_E$ is real, the integrals of interest are now real and no unpleasant
oscillations occur. Moreover, since $S_E$ is bounded from below, the factor
$\exp (-S_E)$ in the integrand is bounded. Strongly fluctuating fields have
a large Euclidean action $S_E$ and are thus suppressed by the factor
$\exp (-S_E)$. (Strictly speaking, this statement does not make sense in
field theory unless renormalization is taken into account.) This makes
Euclidean functional integrals so attractive compared to their Minkowskian
counterparts.

\begin{figure}[hbt]
\begin{center}
\epsfig{file=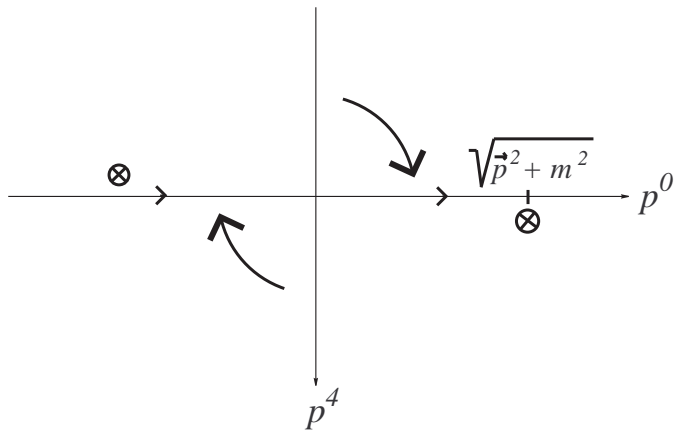,height=6cm}
\caption{Wick rotation in momentum space and the position of the
propagator poles}
\label{wickrot2}
\end{center}
\end{figure}

To illustrate the coordinate transformation to imaginary time,
there is a little exercise.  Consider the Feynman propagator and show that
\begin{equation}
\Delta_F^E(x) = \int\!\frac{d^4p}{(2\pi)^4}\ \frac{e^{ipx}}{p^2+m_0^2},
\end{equation}
(where $px$ is to be understood as a Euclidean scalar product),
is obtained by correct Wick rotation. To be more precise,
\begin{equation}
\Delta_F(\vec x,t) = \lim_{\phi\rightarrow \pi/2}
\Delta_F^E(\vec x, t e^{i\phi}),
\end{equation}
with $\Delta_F$ the Feynman propagator in Minkowski-space
\begin{equation}
\Delta_F(\vec x,t) = i\int\!\frac{d^4p}{(2\pi)^4}\
\frac{e^{-ip*x}}{p^2-m_0^2+i\epsilon},
\end{equation}
where all scalar products in the last expression are defined with Minkowski
metric. An important feature of the Wick-rotated propagator is the absence
of singularities on the $p^4$-axis in Euclidean space, see
Fig.~\ref{wickrot2}.

One might think that in the Euclidean domain everything is unphysical and
there is no possibility to get physical results directly from the Euclidean
Greens functions. But this is not the case. For example, the spectrum of the
theory can be obtained in the following way. Let us consider a vacuum
expectation value of the form
\begin{equation}
\langle 0| A_1 e^{-H\tau} A_2 |0\rangle,
\end{equation}
where the $A_i$'s are formed out of the field $\phi$, e.g.\
$A = \phi(\vec x, 0)$ or $A = \int\!d^3x\ \phi(\vec x, 0)$.
Now, with the familiar insertion of a complete set of energy
eigenstates, we have
\begin{equation}
\langle 0| A_1 e^{-H\tau} A_2 |0\rangle =
\sum_n \langle 0|A_1|n\rangle e^{-E_n\tau} \langle n|A_2|0\rangle .
\end{equation}
In case of a continuous spectrum the sum is to be read as an integral.
On the other hand, representing the expectation value as a functional
integral leads to
\begin{equation}
\frac{1}{Z} \int\!{\cal D}\phi\ e^{-S_E} A_1(\tau)A_2(0) =
\sum_n \langle 0|A_1|n\rangle \langle n|A_2|0\rangle e^{-E_n\tau}.
\end{equation}
This is similar to the ground state projection at the beginning of this
chapter. For large $\tau$ the lowest energy eigenstates will dominate the sum
and we can thus obtain the low-lying spectrum from the asymptotic behaviour
of this expectation value. By choosing $A_1, A_2$ suitably, e.g.\ for
\begin{equation}
A \equiv A_1 = A_2 = \int\!d^3x\ \phi(\vec x,0),
\end{equation}
such that $\langle 0 | A | 1 \rangle \neq 0$ for a one-particle state
$| 1 \rangle$ with zero momentum $\vec p = 0$ and mass $m_1$, we will
get
\begin{equation}
\frac{1}{Z} \int\!{\cal D}\phi\ e^{-S_E} A(\tau) A(0)
= |\langle 0|A|1 \rangle|^2 e^{-m_1\tau} + \dots,
\end{equation}
which means that we can extract the mass of the particle.

{}From now on we shall remain in Euclidean space and suppress the subscript
$E$, so that $S \equiv S_E$ means the Euclidean action.

\subsection{Lattice Discretization}

One central question still remains: does the infinite dimensional
integration over all classical field configurations, i.e.
\begin{equation}
{\cal D}\phi  =  \prod_x d\phi(x),
\end{equation}
make sense at all? How is it defined?

\begin{figure}[hbt]
\begin{center}
\epsfig{file=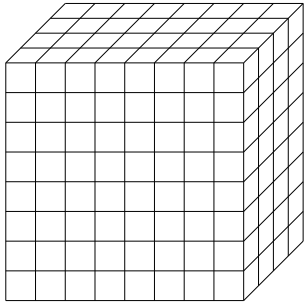,height=4cm}
\caption{3--dimensional lattice}
\label{lattice}
\end{center}
\end{figure}

Remember the way we derived the path integral representation of quantum
mechanics. It was obtained as a limit of a discretization in time $\tau$. As
in field theory the fields depend on the four Euclidean coordinates instead
of a single time coordinate, we may now introduce a discretized space-time
in form of a lattice, for example a hypercubic lattice, specified by
\begin{equation}
x_{\mu} = a n_{\mu}, \qquad n_{\mu}\in {\bf Z},
\end{equation}
see Fig.~\ref{lattice}.

The quantity $a$ is called the lattice spacing for obvious reasons.
The scalar field
\begin{equation}
\phi(x), \qquad x \in \mbox{lattice},
\end{equation}
is now defined on the lattice points only.
Partial derivatives are replaced by finite differences,
\begin{equation}
\partial_{\mu}\phi \longrightarrow \Delta_{\mu}\phi(x) \equiv
\frac{1}{a} (\phi(x+a{\hat{\mu}})-\phi(x)),
\end{equation}
and space-time integrals by sums:
\begin{equation}
\int\!d^4x \quad \longrightarrow  \sum_x a^4 \ .
\end{equation}

The action of our discretized $\phi^4$-theory, Eq.~(\ref{Eaction}),
can be written as
\begin{equation}
S = \sum_x a^4 \left\{ \frac{1}{2} \sum_{\mu =1}^4
(\Delta_{\mu}\phi(x))^2 + \frac{m_0^2}{2}\phi(x)^2 +
\frac{g_0}{4!}\phi(x)^4
\right\}.
\end{equation}

In the functional integrals the measure
\begin{equation}
{\cal D}\phi  =  \prod_x d\phi(x)
\end{equation}
involves the lattice points $x$ only. So we have a discrete set
of variables to integrate. If the lattice is taken to be finite, we just
have finite dimensional integrals.

Discretization of space-time using lattices has one very important
consequence. Due to a non-zero lattice spacing a cutoff in momentum space
arises. The cutoff can be observed by having a look at the Fourier
transformed field
\begin{equation}
\tilde{\phi}(p) = \sum_x a^4\ e^{-ipx}\ \phi(x).
\end{equation}
The Fourier transformed functions are periodic in momentum-space, so that we
can identify
\begin{equation}
p_{\mu} \cong p_{\mu}+\frac{2\pi}{a}
\end{equation}
and restrict the momenta to the so-called Brillouin zone
\begin{equation}
-\frac{\pi}{a} \, < \, p_{\mu}\,\leq \frac{\pi}{a}.
\end{equation}
The inverse Fourier transformation, for example, is given by
\begin{equation}
\phi(x) = \int_{-\pi/a}^{\pi/a} \frac{d^4p}{(2\pi)^4}\ e^{ipx}\
\tilde{\phi}(p).
\end{equation}
We recognize an ultraviolet cutoff
\begin{equation}
|p_{\mu}| \leq \frac{\pi}{a}.
\end{equation}
Therefore field theories on a lattice are regularized in a natural way.

In order to begin in a well-defined way one would start with a finite
lattice. Let us assume a hypercubic lattice with length $L_1=L_2=L_3=L$ in
every spatial direction and length $L_4=T$ in Euclidean time,
\begin{equation}
x_{\mu} = an_{\mu},\qquad n_{\mu} = 0,1,2,\dots,L_{\mu}-1,
\end{equation}
with finite volume $V = L^3T$. In a finite volume one has to specify boundary
conditions. A popular choice are periodic boundary conditions
\begin{equation}
\phi(x) = \phi(x+aL_{\mu}\,\hat{\mu}),
\end{equation}
where $\hat{\mu}$ is the unit vector in the $\mu$-direction.
They imply that the momenta are also discretized,
\begin{equation}
p_{\mu} = \frac{2\pi}{a}\,\frac{l_{\mu}}{L_{\mu}}\qquad \mbox{with}\
l_{\mu} = 0,1,2,\dots,L_{\mu}-1,
\end{equation}
and therefore momentum-space integration is replaced by finite sums
\begin{equation}
\int\!\frac{d^4p}{(2\pi)^4}\ \longrightarrow \
\frac{1}{a^4 L^3 T}\sum_{l_{\mu}}.
\end{equation}
Now, all functional integrals have turned into regularized and finite
expressions.

Of course, one would like to recover physics in a continuous and infinite
space-time eventually. The task is therefore to take the infinite volume
limit,
\begin{equation}
L,T \longrightarrow \infty,
\end{equation}
which is the easier part in general, and to take the the
continuum limit,
\begin{equation}
a \longrightarrow  0.
\end{equation}
Constructing the continuum limit of a lattice field theory is usually
highly nontrivial and most effort is often spent here.

The formulation of Euclidean quantum field theory on a lattice bears a
useful analogy to statistical mechanics. Functional integrals have the form
of partition functions and we can set up the following correspondence:

\begin{center}
\begin{tabular}{|c|c|}\hline
Euclidean field theory & Statistical Mechanics\\
\hline
generating functional & partition function \\
$\int\!{\cal D}\phi\ e^{-S}$ & $\sum e^{-\beta {\cal H }}$ \\
\hline
action & Hamilton function \\
$S$ & $\beta {\cal H}$ \\
\hline
mass $m$ & inverse correlation length $1 / \xi$\\
$G \sim e^{-mt}$ & $G \sim e^{-\frac{x}{\xi}}$\\
\hline
\end{tabular}
\end{center}
This formal analogy allows to use well established methods of statistical
mechanics in field theory and vice versa.  Even the terminology of both
fields is often identical. To mention some examples, in field theory one
employs high-temperature expansions and mean field approximations, and in
statistical mechanics one applies the renormalization group.

\section{Lattice Gauge Theory}

In this section gauge fields are implemented on a space-time lattice. After
introducing the concept of parallel transporters we define the variables of
lattice gauge theory and consider some aspects of pure gauge theory,
including static confinement and the glueball spectrum.

\subsection{Parallel Transporters}

Let us start with a brief reminder of gauge transformations in continuum
field theory. For an $N$-component complex scalar field
$\phi(x) = (\phi^i(x)),\quad i=1,\dots,N$, gauge transformations are defined
as
\begin{equation}
\phi(x) \longrightarrow \Lambda(x) \phi (x),\qquad
\mbox{with}\ \Lambda (x) \in {\rm SU}(N),
\end{equation}
where we consider the case of gauge group SU($N$).

In order to restore the invariance of the Lagrangian under these
transformations, one introduces a covariant derivative
\begin{equation}
D_{\mu}\phi(x) = (\partial_{\mu}-ig_0 A_{\mu}^a(x) T_a)\,\phi(x),
\end{equation}
where the $T^a$'s are generators of the gauge group
and $A_{\mu}^a(x)$ is the gauge field.
For SU($N$) there are $(N^2-1)$ generators $T_a$,
$a = 1,2,3,\dots, N^2-1$, and they satisfy
\begin{equation}
\lbrack T_a,T_b \rbrack = if_{abc}T_c
\end{equation}
with the structure constants $f_{abc}$ of SU($N$).
For SU(2) the three generators are given by the Pauli matrices
\begin{equation}
T_a =  \frac{\sigma_a}{2},\qquad a=1,2,3,
\end{equation}
and for SU(3) they are the Gell-Mann matrices
\begin{equation}
T_a =  \frac{\lambda_a}{2},\qquad a=1,\dots,8.
\end{equation}

It is now an easy exercise to show that the covariant derivative of
a scalar field transforms covariantly under the gauge transformation:
\begin{equation}
D_{\mu}\phi(x) \longrightarrow  \Lambda(x) D_{\mu}\phi(x),
\end{equation}
and therefore a kinetic term $D_{\mu}\phi(x) \cdot D_{\mu}\phi(x)$
is invariant under gauge transformations.

Associated with the gauge field is its field strength
\begin{equation}
F_{\mu \nu}^a(x) = \partial_{\mu} A_{\nu}^a
-\partial_{\nu} A_{\mu}^a + g_0 f_{abc} A_{\mu}^b A_{\nu}^c.
\end{equation}
With the help of it an action for the gauge field, the Yang-Mills
action, can be written down as
\begin{equation}
S_{YM} = \frac{1}{4}\int\!d^4x\ F_{\mu \nu}^a(x) F_{\mu \nu}^a(x).
\end{equation}
Now let us consider a product of fields at different points,
\begin{equation}
\phi(x) \cdot \phi(y) = \sum_i \overline{\phi^i(x)} \, \phi^i(y).
\end{equation}
This is not invariant under gauge transformations. But such terms will occur,
if we write down a kinetic term in lattice field theory. Therefore, we need
matrices $U(x,y) \in {\rm SU}(N)$ which transform as
$U(x,y) \longrightarrow \Lambda(x) U(x,y) \Lambda^{-1}(y)$, such that
$\phi(x) \cdot U(x,y) \phi(y)$ would be invariant. There is a solution to
this problem. Take a path $\cal C$ from $y$ to $x$ and define
\begin{equation}
U(x,y;{\cal C}) \equiv P \exp{ig_0 \int_y^x A_{\mu}^a(z)T_a dz^{\mu}},
\end{equation}
where the integral is taken along the path $\cal C$ and the symbol $P$
indicates a path ordering of the non-commuting factors $A_{\mu}^a(z)T_a$
like the time ordering in Dyson's formula.

\begin{figure}[hbt]
\begin{center}
\epsfig{file=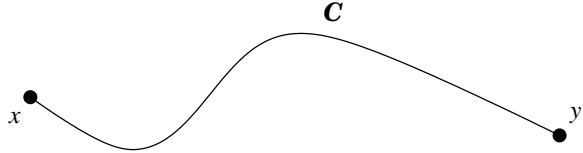,height=2cm}
\caption{Path ${\cal C}$ between $y$ and $x$}
\label{paralleltransport}
\end{center}
\end{figure}

Then $U(x,y;{\cal C})$ transforms as desired and fulfills the goal. It is
called {\em parallel transporter} in analogy to similar objects in
differential geometry, which map vectors from one point to another along
curves. The parallel transporters depend not only on the points $x$ and $y$
but also on the chosen curve $\cal C$. They obey the composition rule
\begin{equation}
U(x,y;{\cal C}) = U(x,u;{\cal C}_1) \cdot U(u,y;{\cal C}_2),
\end{equation}
where the path $\cal C$ is split into two parts ${\cal C}_1$ and
${\cal C}_2$.

In the Abelian case, where the gauge group is U(1), the path ordering
$P$ is not required and we just have
\begin{equation}
U(x,y;{\cal C}) =  \exp{ig_0 \int_y^x A_{\mu}(z)dz^{\mu}}.
\end{equation}

\subsection{Lattice Gauge Fields}

The kinetic term for scalar fields on a lattice involves the product of
fields at neighbouring lattice points, separated by the lattice spacing
$a$. In order to make it gauge invariant we need the smallest parallel
transporters, which exist on a lattice, namely those connecting nearest
neighbour points. The corresponding paths are called {\em links}.

\begin{figure}[hbt]
\begin{center}
\epsfig{file=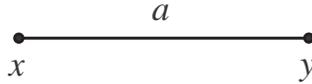,height=2cm}
\caption{Link $b = \langle x,y \rangle$ between lattice points $x$ and
$y$}
\label{link}
\end{center}
\end{figure}

\begin{figure}[hbt]
\begin{center}
\epsfig{file=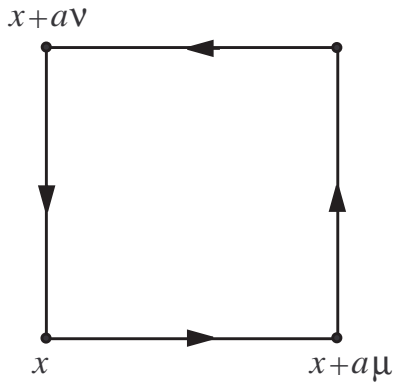,height=6cm}
\caption{Plaquette $p$ in lattice directions $\mu$ and $\nu$}
\label{plaquette}
\end{center}
\end{figure}

With each link
\begin{equation}
b = \langle x+a\hat{\mu},x \rangle
\end{equation}
in lattice direction $\hat{\mu}$ we associate a sort of elementary parallel
transporter, the link variable
\begin{equation}
U(b) \equiv U(x+a\hat{\mu},x) \equiv U_{x \mu} \in {\rm SU}(N).
\end{equation}
They replace the gauge field $A_{\mu}^a(x)$. The link variables transform as
\begin{equation}
U(x,y) \longrightarrow \Lambda(x) U(x,y) \Lambda^{-1}(y)
\end{equation}
and therefore in the action a term of the form
\begin{equation}
\sum_{x,\mu} \phi(x+a\hat{\mu}) \cdot U_{x \mu}\, \phi(x)
\end{equation}
is invariant. There exist other gauge invariant expressions on the lattice.
Particularly interesting is the trace of a product of link variables along a
closed path,
\begin{equation}
{\rm Tr}(U(b_1)U(b_2)\dots U(b_n)).
\end{equation}
The most elementary one is a plaquette, as shown in Fig.~\ref{plaquette}.

The plaquette variable
\begin{equation}
U(p) = U_{x\mu \nu} \equiv
U_{(x+a\hat{\nu}) (-\nu)}
U_{(x+a\hat{\mu}+a\hat{\nu})(-\mu)} 
U_{(x+a\hat{\mu}) \nu}
U_{x \mu} 
\end{equation}
has been used by Wilson to construct a lattice Yang-Mills action.
He proposed
\begin{equation}
\label{Wilsonaction}
S_W = -\sum_p \frac{\beta}{N} {\rm Re}({\rm Tr}(U(p))),
\end{equation}
which is gauge invariant and purely real by construction.  The constant
$\beta$ is to be determined in the continuum limit requiring that the
standard Yang-Mills action is recovered in that case. If we introduce gauge
field variables by
\begin{equation}
U_{x \mu} \equiv \exp{ig_0 a A_{\mu}^b(x)T_b}
\end{equation}
then we obtain in a naive continuum limit, where $a$ goes to zero,
\begin{equation}
S_W = \frac{\beta g_0 ^2}{8N}\sum_x a^4 F_{\mu \nu}^b F_{\mu \nu}^b
+ {\cal O}(a^5),
\end{equation}
and we can read off
\begin{equation}
\beta = \frac{2N}{g_0^2}.
\end{equation}

For the quantum theory we have to specify how to do functional integrals.
The integral over all gauge field configurations on the lattice amounts to
an integral over all link variables $U(b)$. So, for the expectation value of
any observable $A$ we write
\begin{equation}
\langle A \rangle = \frac{1}{Z}\int\!\prod_b dU(b)\ A\ e^{-S_W},
\end{equation}
where the integration $dU(b)$ for a given link $b$ is to be understood
as the invariant integration over the group manifold (e.g.\ a 3-sphere
for SU(2)), normalized to
\begin{equation}
\int\!dU = 1.
\end{equation}
As a shorthand, we shall write
\begin{equation}
{\cal D}U \equiv \prod_b dU(b).
\end{equation}
It is worth noticing here that no gauge fixing appears to be necessary.
The total ``volume of the gauge group'' is unity. Gauge fixing is
required for the purpose of perturbation theory only.

\subsection{Some Observables}

Already in pure gauge theory some quantities of physical interest
occur and we shall consider two of them.

\begin{figure}[hbt]
\begin{center}
\epsfig{file=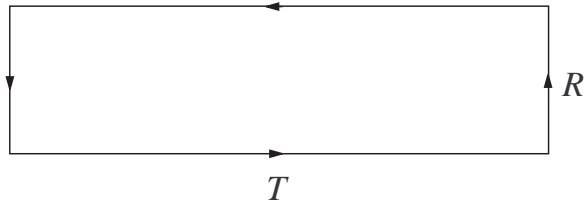,width=8cm}
\caption{Wilson loop}
\label{Wloop}
\end{center}
\end{figure}

1. The Wilson loop is defined as the trace of a parallel transporter for
a closed curve ${\cal C}$ of spatial length $R$ and time extension $T$,
see Fig.~\ref{Wloop}.

In the limit $T\rightarrow \infty$ it can be shown that
\begin{equation}
\langle {\rm Tr}(U({\cal C}))\rangle \sim C\exp{-TV(R)},
\end{equation}
with $V(R)$ being the potential between static colour charges, the
so-called static quark-antiquark potential.

If for large areas $RT$ the Wilson loop goes to zero with an area law
of the form
\begin{equation}
\langle {\rm Tr}(U({\cal C}))\rangle \sim \exp{-\alpha R T},
\end{equation}
the potential rises linearly for large $R$:
\begin{equation}
V(R)\,\sim\,\alpha R.
\end{equation}
Such a situation is called static quark confinement, since colour
charges feel a constant attractive force at arbitrary large distances.

\begin{figure}[hbt]
\begin{center}
\epsfig{file=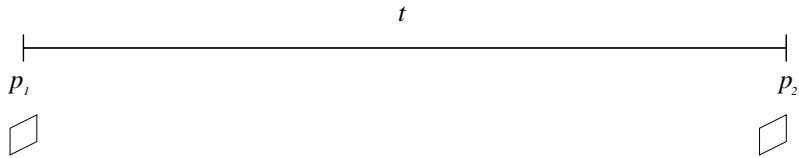,height=2cm}
\caption{Plaquette--plaquette correlations}
\label{plaquettecor}
\end{center}
\end{figure}

2. Plaquette correlations are expectation values of the product of two
spatial plaquettes $p_1$ and $p_2$, separated by a time $t$ as
illustrated in Fig.~\ref{plaquettecor}.

In non-abelian lattice gauge theories one finds that these correlations fall
off exponentially according to
\begin{equation}
\langle {\rm Tr}(U(p_1)){\rm Tr}(U(p_2)) \rangle_c \sim \exp{-mt}.
\end{equation}
{}From our general discussion it follows that $m$ is the lowest particle
mass in the theory. Since there are only gluonic degrees of freedom present,
the corresponding massive particle is called {\em glueball}.

\section{Methods}

In the previous section the functional integrals for gauge theories on the
lattice have been defined. But it is another problem to evaluate these
highdimensional integrals. A calculation in closed form appears to be
impossible in general. In this section we shall consider some of the methods
used to evaluate the functional integrals approximately.

\subsection{Perturbation Theory}

Although lattice gauge theory offers us the possibility to study
non-per\-tur\-ba\-tive aspects, perturbation theory is nevertheless a highly
valuable tool on the lattice, too. In particular, it is used to relate
perturbatively and non-per\-tur\-ba\-tive\-ly calculated quantities.

Perturbation theory amounts to an expansion in powers of the coupling as in
the continuum. The lattice provides an intrinsic UV cutoff $\pi / a$ for all
momenta. Apart from that one has to observe that the propagators and
vertices are different from the continuum ones, owing to the form of the
lattice action. In particular, gluon self interactions of all orders appear
and not only as three and four gluon vertices. As mentioned before,
perturbation theory on the lattice requires gauge fixing as is the case in
the continuum.

\subsection{Strong Coupling Expansion}

We have already pointed out the analogies between Euclidean field theory and
statistical mechanics. In statistical mechanics a well-established technique
is the high-temperature expansion. This is an expansion in powers of
\begin{equation}
\beta \sim \frac{1}{g_0^2},
\end{equation}
which is a small quantity at large bare couplings $g_0$. Therefore it is
the same as a strong coupling expansion. Basically the Boltzmann factor
is expanded as
\begin{equation}
\exp{\beta\frac{1}{N} {\rm Re}({\rm Tr}(U(p)))}
= 1 + \beta \frac{1}{N} {\rm Re}({\rm Tr}(U(p))) + \dots\ .
\end{equation}
The resulting expansion can be represented diagrammatically, similar to the
Feynman diagrams of perturbation theory. The diagram elements, however, are
plaquettes $p$ on the lattice. Every power of $\beta$ introduces one more
plaquette.

\begin{figure}[hbt]
\begin{center}
\epsfig{file=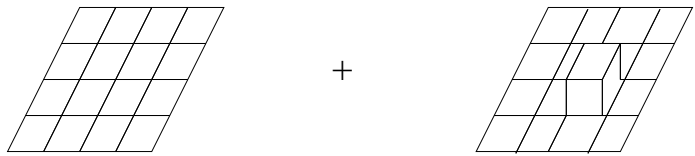,height=2cm}
\caption{Leading terms in the diagrammatic representation of the strong
coupling expansion of a Wilson loop}
\label{plaquetteexp}
\end{center}
\end{figure}

For example, the calculation of a Wilson loop in the strong coupling
expansion leads to surfaces that are bounded by the loop $\cal C$, see
Fig.~\ref{plaquetteexp}.

These diagrams lead to an area law of the form
\begin{equation}
\langle {\rm Tr} (U({\cal C})) \rangle \sim e^{-\alpha RT} \qquad
\mbox{with} \quad \alpha = - \ln \beta + \dots = \ln g_0^2 + \dots .
\end{equation}

\begin{figure}[hbt]
\begin{center}
\epsfig{file=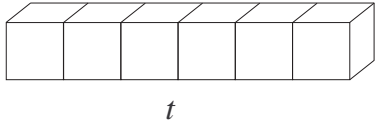,height=2cm}
\caption{Leading term in the diagrammatic representation of the strong
coupling expansion of a plaquette correlation}
\label{correlationexp}
\end{center}
\end{figure}

The strong coupling expansion of plaquette correlations is associated with
surfaces connecting the two plaquettes. The smallest of them is a tube of
cross section 1 between the plaquettes, with total area $4t$, see
Fig.~\ref{correlationexp}.

Summing these diagrams gives
\begin{equation}
\langle {\rm Tr}(U(p_1)){\rm Tr}(U(p_2)) \rangle_c \sim e^{-mt}
\qquad \mbox{with} \quad m = -4 \ln \beta + \dots .
\end{equation}
This is a general result for lattice gauge theories: for small $\beta$,
i.e.\ for large bare couplings $g_0$, there is static confinement,
$\alpha > 0$, and dynamical mass generation, $m>0$. These are genuine
non-perturbative properties.

\subsection{Other Analytic Methods}

Other analytical methods are available for approximative evaluations of the
functional integrals of lattice gauge theory. Instead of going into the
details we just want to mention some of them:
\begin{itemize}
\item mean field approximation
\item renormalization group
\item $\frac{1}{N}$-expansion
\end{itemize}

\subsection{Monte Carlo Methods}

\begin{figure}[hbt]
\begin{center}
\epsfig{file=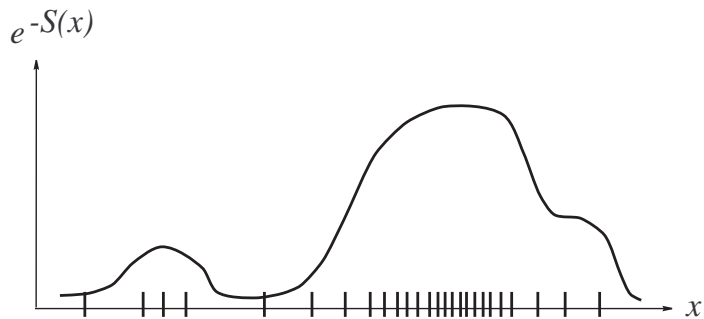,width=10cm}
\caption{One-dimensional distribution of importance sampled quadrature
points}
\label{importancesampling}
\end{center}
\end{figure}

On a finite lattice the calculation of expectation values requires the
evaluation of finite dimensional integrals. This immediately suggests the
application of numerical methods. The first thing one would naively propose
is some simple numerical quadrature.  In order to understand that this
approach wouldn't be all that helpful, let us consider a typical lattice as
it is considered in recent calculations.  With 40 lattice points in every
direction we have $4 \cdot 40^4$ link variables. For gauge group SU(3) this
gives 81,920,000 real variables. That should be intractable for conventional
quadratures even in the future. Therefore some statistical method is
required. Producing lattice gauge configurations just randomly turns out to
be extremely inefficient. The crucial idea to handle this problem is the
concept of {\em importance sampling}: for a given lattice action $S$
quadrature points $x_i$ are generated with a probability
\begin{equation}
p(x_i) \sim \exp{-S(x_i)}.
\end{equation}
This provides us with a large number of points in the important regions
of the integral, improving the accuracy drastically.
See Fig.~\ref{importancesampling} for a one-dimensional sketch.

In case of lattice gauge theory the quadrature points are configurations
$U^{(i)} = \left\{ U_{x\mu}^{(i)} \right\}$. An expectation value
\begin{equation}
\langle 0 | A | 0 \rangle = \frac{1}{Z} \int\!{\cal D}U\ A(U)\
e^{-S(U)}
\end{equation}
is numerically approximated by the average
\begin{equation}
\bar{A} \equiv \frac{1}{n} \sum_{i=1}^n A( U^{(i)} ).
\end{equation}

The Monte Carlo method consists in producing a sequence of configurations
$U^{(1)} \rightarrow U^{(2)} \rightarrow U^{(3)} \rightarrow \dots$ with the
appropriate probabilities in a statistical way. This is of course done on a
computer. An {\em update} is a step where a single link variable $U_{x\mu}$
is changed, whereas a {\em sweep} implies that one goes once through the
entire lattice, updating all link variables. A commonly used technique for
obtaining updates is the {\em Metropolis algorithm}.

An important feature of this statistical way of evaluation is the existence
of statistical errors.  The result of such a calculation is usually
presented in the form
\begin{equation}
\langle A \rangle = \bar{A} \pm \sigma_{\bar{A}},
\end{equation}
where the variance of $\bar{A}$ decreases with the number $n$ of
configurations as
\begin{equation}
\sigma_{\bar{A}} \sim \frac{1}{n^{1/2}}.
\end{equation}

\section{Fermions on the Lattice}

Maybe the reader is already asking themself:  when do they finally start
discussing QCD?  To approach the strong interactions, an implementation of
quark fields on the lattice is necessary.  We shall demonstrate the
foundations, techniques and difficulties of that task reviewing the standard
representation of fermions with Grassmann variables and discussing several
ansaetze to discretize this continuum description. This will lead us to the
question of fermion doubling on the lattice - a special problem in the
chiral limit.

\subsection{Grassmann Variables}

First remember scalar fields in the continuum.
Classical fields are just ordinary functions and satisfy
\begin{equation}
[\phi(x),\phi(y)] = 0,
\end{equation}
which can be considered as the limit $\hbar \rightarrow 0$ of the quantum
commutation relations.

Fermi statistics implies that fermionic quantum fields have the well-known
equal-time anticommutation relations
\begin{equation}
\{\psi(\vec x,t),\psi(\vec y,t)\} = 0.
\end{equation}
Motivated by this, we might introduce a classical limit in which classical
fermionic fields satisfy
\begin{equation}
\{\psi(x),\psi(y)\}\,=0
\end{equation}
for all $x,y$.
Classical fermionic fields are therefore anticommuting variables, which are
also called {\em Grassmann variables}.

We would like to point out that the argument above is just a heuristic
motivation. More rigorous approaches can be found in the literature.

In general, a complex Grassmann algebra is generated by elements $\eta_i$
and $\bar{\eta}_i$, which obey
\begin{eqnarray}
\{\eta_i,\eta_j\} &=& 0 \\
\{\eta_i,\bar{\eta}_j\} &=& 0 \\
\{\bar{\eta}_i,\bar{\eta}_j\} &=& 0.
\end{eqnarray}
An integration of Grassmann variables can be defined by
\begin{equation}
\int\!d\eta_i\ (a+b\eta_i) = b
\end{equation}
for arbitrary complex numbers $a,b$.

In fermionic field theories we have Grassmann fields, which associate
Grassmann variables whith every space-time point. For example, a Dirac field
has anticommuting variables $\psi_{\alpha}(x)$ and $\bar{\psi}_{\alpha}(x)$,
where $\alpha$=1,2,3,4 is the Dirac index. The classical Dirac field obeys
\begin{equation}
\{ \psi_{\alpha}(x), \psi_{\beta}(y) \} = 0, \quad \mbox{etc.}\,.
\end{equation}
In order to write down fermionic path integrals as integrals over
fermionic and anti-fermionic field configurations, we write
\begin{equation}
{\cal D}\psi\, {\cal D}\bar{\psi} = \prod_x \prod_{\alpha}
d\psi_{\alpha}(x)\, d\bar{\psi}_{\alpha}(x).
\end{equation}
Then any fermionic Greens function is of the form
\begin{equation}
\langle 0|A|0\rangle  = \frac{1}{Z}\int\!{\cal D}\psi\,{\cal D}\bar{\psi}\
A\ e^{-S_F},
\end{equation}
with an action $S_F$ for the fermions. For a free Dirac field the action is
\begin{equation}
S_F = \int\!d^4x\ \bar{\psi}(x) (\gamma_{\mu}\partial^{\mu}+m)\psi(x).
\end{equation}
In the context of the standard model, fermionic actions are always bilinear
in the fermionic fields. With the help of the Grassmann integration rules
above one can then show that the functional integrals are formally
remarkably simple to calculate:
\begin{equation}
\int\!{\cal D}\psi{\cal D}\bar{\psi}\ e^{-\int\!d^4x\,
\bar{\Psi}(x) Q \Psi(x)} = \det{Q}.
\label{detQ}
\end{equation}
This is the famous fermion determinant. The main problem remains, of course,
namely to evaluate the determinant of the typically huge matrix $Q$.

\subsection{Naive Fermions}

So far no difficulties for the implementation of fermions on the lattice
seem to arise: all we have to do is to discretize the field configurations
in the well-known way and to calculate the Greens functions with some of the
methods of the last section. However, we fail.  To see this, let us consider
the propagator of a fermion with mass $m$ as an example.  The fermionic
lattice action is then given by
\begin{equation}
S_F = \frac{1}{2} \sum_x \sum_{\mu} \bar{\psi}(x) (\gamma_{\mu}
\Delta_{\mu} + m)\psi(x) + h.c.
\end{equation}
and the resulting propagator is
\begin{equation}
\tilde{\Delta}(k) = \frac{-i\sum_{\mu}\gamma_{\mu}\sin{k_{\mu}}+m}
{\sum_{\mu}\sin{k_{\mu}}^2+m^2}.
\end{equation}
The propagator has got a pole for small $k_{\mu}$ representing the physical
particle, but there are additional poles near $k_{\mu} = \pm \pi$ due to the
periodicity of the denominator. So $S_F$ really describes 16 instead of 1
particle. This problem - euphemistically called fermion doubling - is a
crucial obstacle for all lattice representations of quark fields and we
shall next discuss some of the possible ways out of this dilemma.

\subsection{Wilson vs.\ Staggered Fermions}

Fermion doubling was already known to Wilson in the early days of lattice
QCD.  He proposed a modified action for the fermions in order to damp out
the doubled fields in the continuum limit.  Therefore he added another term,
the Wilson term, to the naive action.
\begin{eqnarray}
S_F \rightarrow S_F^{(W)} &=& S_F - \frac{r}{2}\sum_x
\bar{\psi}(x) \Box \psi(x) \nonumber\\
&=& S_F - \frac{r}{2} \sum_{x,\mu} \bar{\psi}(x) \{
\psi(x+\hat{\mu}) + \psi(x-\hat{\mu}) - 2 \psi(x) \},
\end{eqnarray}
where $0< r \le 1$. Calculating the propagator with this modified action,
one finds that the unwanted doubled fermions acquire masses $\propto 1 / a$,
so that they become infinitely massive in the continuum limit and dissapear
from the physical spectrum.

Wilson fermions have a serious disadvantage:  even at vanishing fermion
masses, chiral symmetry is broken explicitely by the Wilson term, and one
has problems with calculations for which chiral symmetry is of central
importance.

There are alternatives to Wilson's approach. One of them, due to Kogut and
Susskind, are so-called staggered fermions. The idea is to distribute the
components $\psi_{\alpha}$ of the Dirac field on different lattice points.
It results in a reduction from 16 to 4 fermions. Moreover, for massless
fermions a remnant of chiral symmetry in form of a chiral
U(1)$\otimes$U(1)-symmetry remains.

\subsection{QCD on the Lattice}

Neglecting the problem of chiral symmetry for the moment, we are now
able to write down QCD for a discretized space-time. Take Wilson's
or staggered fermions for the quark fields
\begin{equation}
\psi^{f}_{c\alpha}(x),
\end{equation}
where $f=1,\dots,N_f$ and $c=1,2,3$ are flavour and colour indices,
and define the action as a sum of the pure gauge action for SU(3),
\begin{equation}
S_G = \beta \sum_p \left( 1 - \frac{1}{3}{\rm Re}{\rm Tr}U(p)\right),
\end{equation}
and the fermionic part $S_F$:
\begin{equation}
S = S_F+S_G.
\end{equation}
Since computers cannot handle anticommuting numbers, in Monte Carlo
calculations one performs the integration over the fermionic fields
explicitly as in Eq.~(\ref{detQ}) and works with an effective action for the
bosonic fields,
\begin{equation}
e^{-S_{eff}(U)} \equiv e^{-S_G(U)} \cdot \det{Q}(U),
\end{equation}
which involves the fermion determinant. Because the calculation of $\det{Q}$
turns out to be very tedious, one often uses the quenched approximation that
treats $Q$ as a constant. In recent years different unquenched
investigations of QCD have been made and have given estimates for quenching
errors.

\section{Physics}

After having introduced the concepts and some of the technicalities, it is
now time to turn to physics and to discuss physically relevant results. For
this purpose it is necessary to talk about the way the continuum limit
should be taken.

\subsection{Continuum Limit}

As we are only able to perform calculations at finite lattice spacing, it is
an important issue to get the extrapolation process to the continuum limit
under control. Since the lattice spacing is the regulator of our theory, it
should be useful to apply renormalization group techniques to this problem. 
Knowing the functional dependence of the bare coupling $g_0$ on the
regulator, in other words solving the renormalization group equation, we
should know how to vary the bare coupling of our theory in order to reach a
continuum limit. Let us discuss this idea in more detail.

\begin{figure}[hbt]
\begin{center}
\epsfig{file=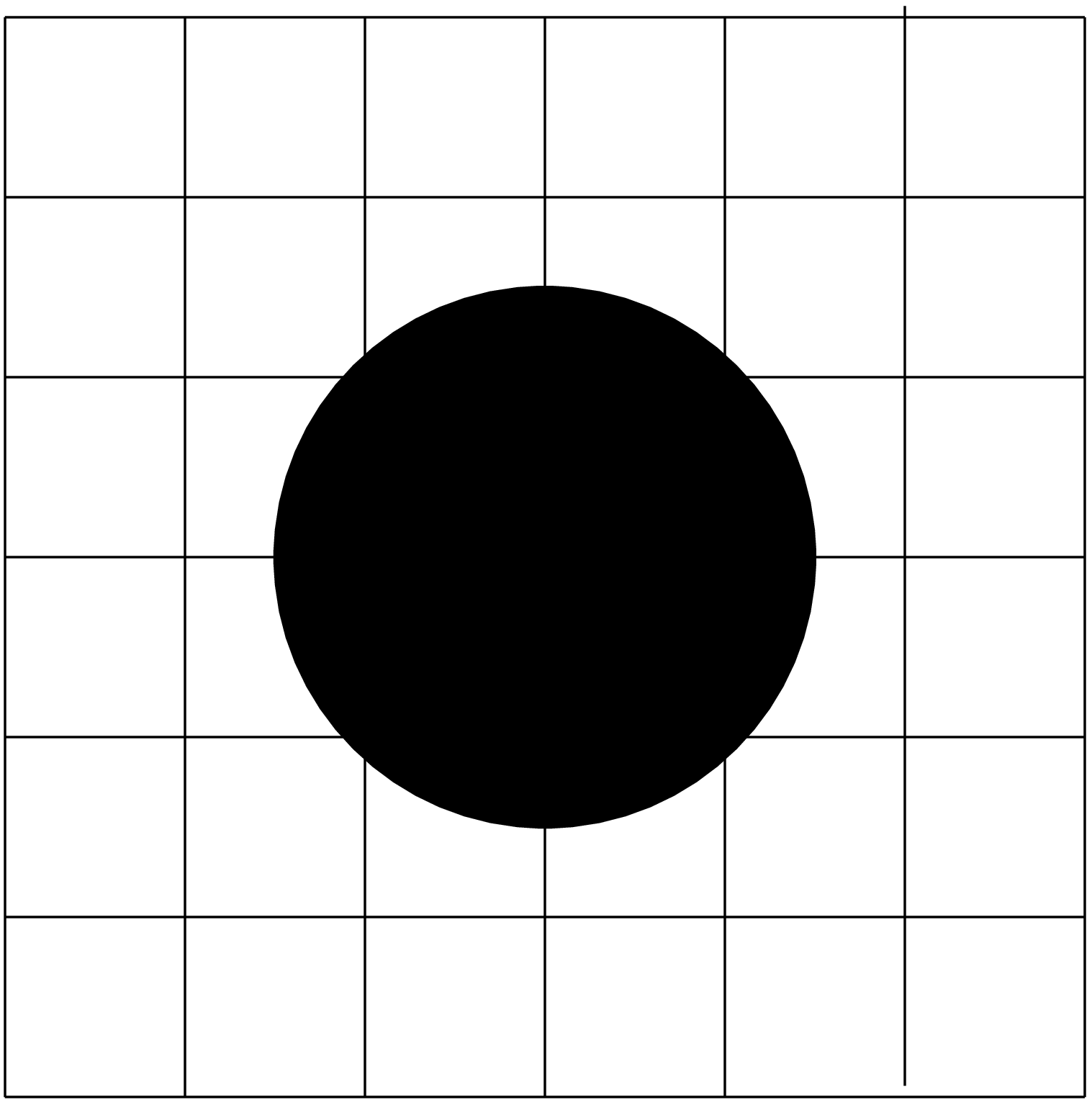,height=3cm}
\epsfig{file=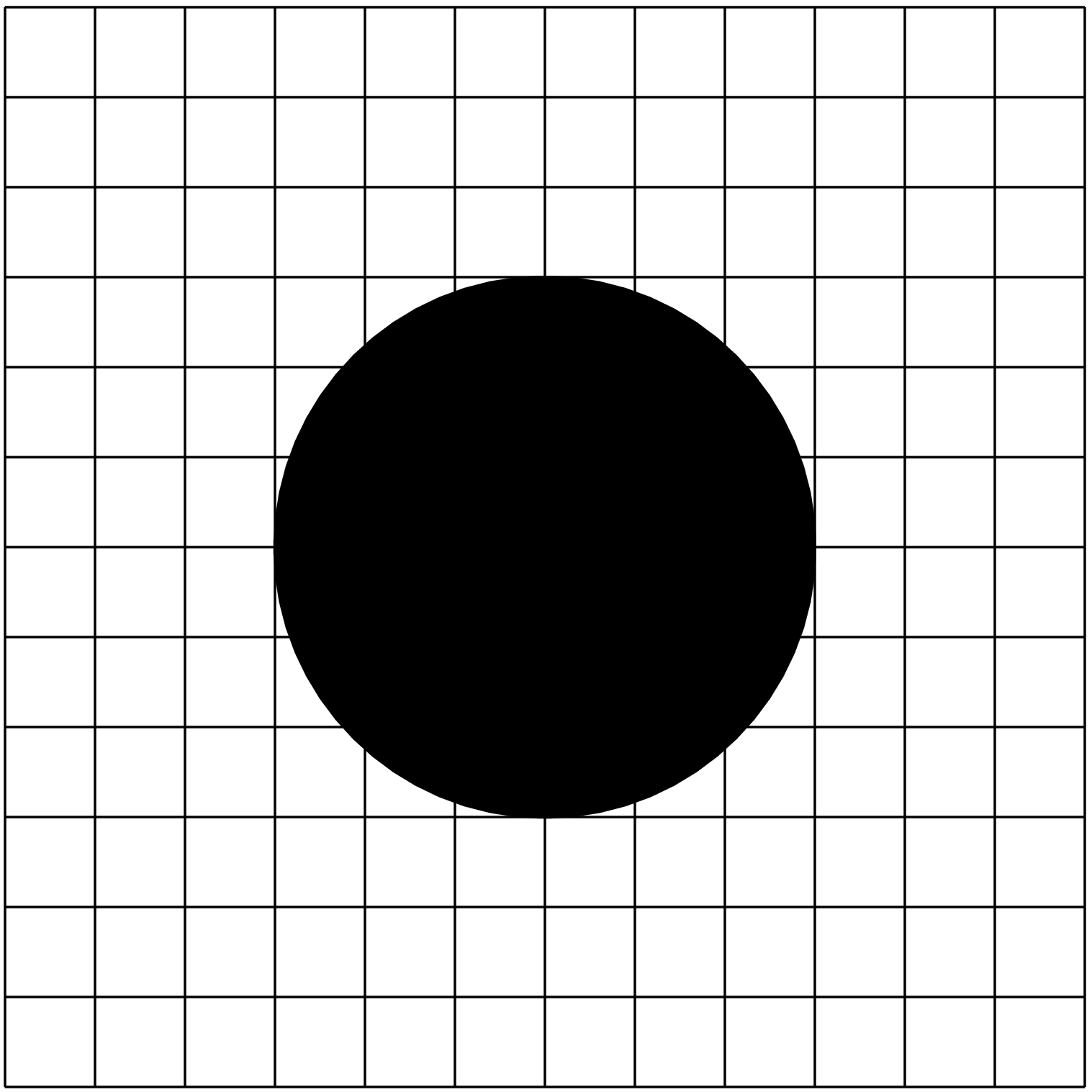,height=3cm}
\epsfig{file=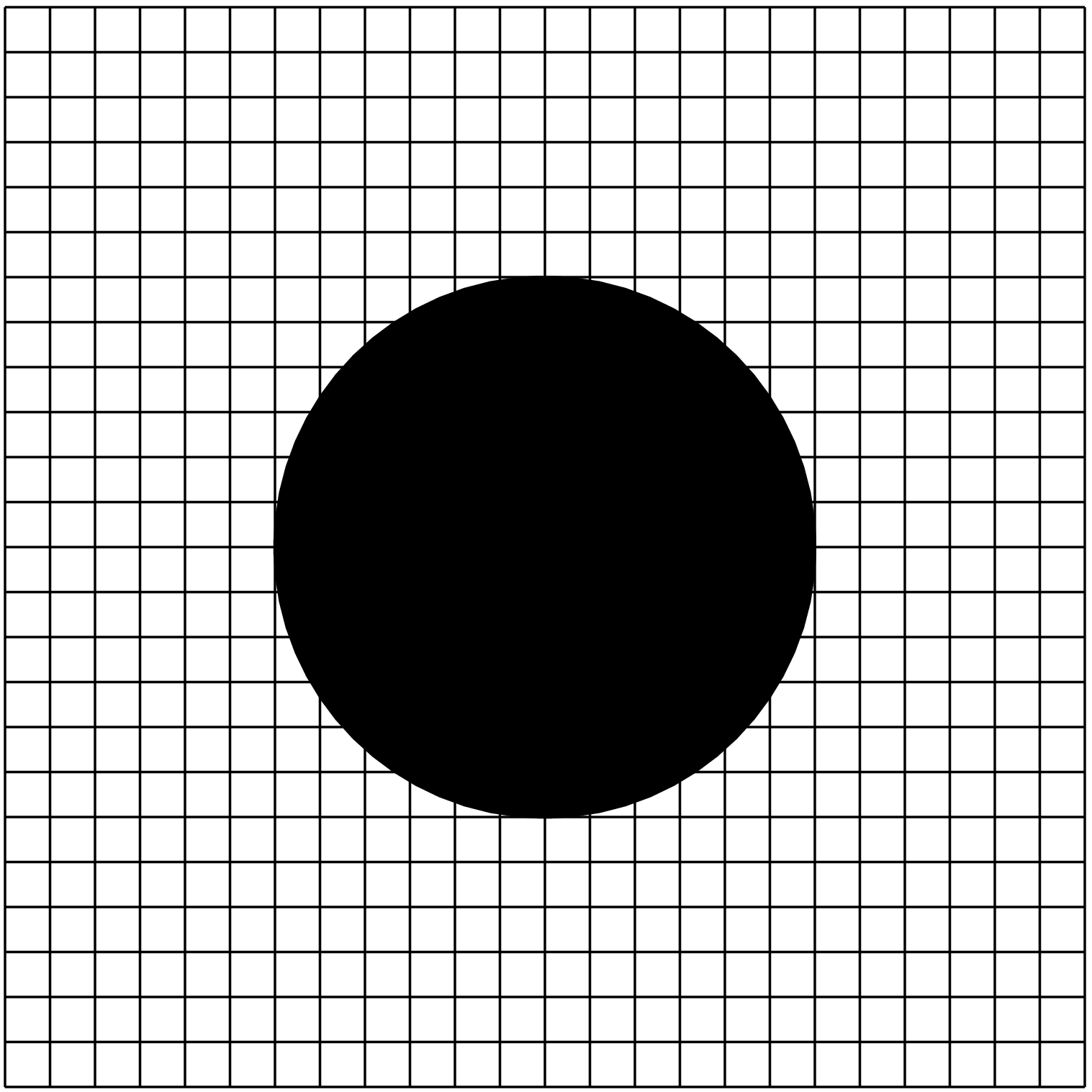,height=3cm}
\caption{2-dimensional lattices with increasing correlation lengths $\xi$}
\label{spacing}
\end{center}
\end{figure}

In the continuum limit the lattice spacing $a$ is supposed to go to zero,
while physical masses $m$ should approach a finite limit. The lattice
spacing, however, is not a dimensionless quantity, therefore we have to fix
some mass scale $m$, e.g.\ some particle mass, and consider the limit
$a m \rightarrow 0$. The inverse of that,
\begin{equation}
\frac{1}{am} \equiv \xi,
\end{equation}
can be regarded as a correlation length. In the continuum limit $\xi$
has to go to infinity, which is called a {\em critical point} of the theory.
In Fig.~\ref{spacing} this is illustrated on a two-dimensional lattice
with different correlation lengths.

In pure gauge theory, there is a single, dimensionless bare coupling $g_0$
and $am$ is clearly a function of $g_0$. In order to approach the
continuum limit, we have to vary $g_0$ such that $am \rightarrow 0$. How
this is done, is controlled by a renormalization group equation:
\begin{equation}
-a \frac{\partial g_0}{\partial a} = \beta_{LAT}(g_0) =
-\beta_0 g_0^3 - \beta_1 g_0^5 + \dots,
\end{equation}
where the first term of the expansion is
\begin{equation}
\beta_0 = \frac{11}{3}\, N\, \frac{1}{16\pi^2}.
\end{equation}
In the perturbative regime of $g_0$ this equation implies that for
decreasing $am$ the bare coupling $g_0$ is also decreasing, getting even
closer to zero. Hence the continuum limit is associated with the limit
\begin{equation}
g_0 \rightarrow 0 \qquad \mbox{(continuum limit).} 
\end{equation}
The solution of the renormalization group equation up to second order in
$g_0$ is
\begin{equation}
a = \Lambda^{-1}_{LAT}\ \exp \left(-\frac{1}{2\beta_0 g_0^2}\right)\
(\beta_0 g_0^2)^{-\frac{\beta_1}{2\beta_0^2}}\ \{1+{\cal O}(g_0^2) \},
\end{equation}
where the lattice $\Lambda$-parameter $\Lambda_{LAT}$ appears.
Solving for $g_0$ yields
\begin{equation}
g_0^2 = \frac{-1}{\beta_0 \log{a^2\Lambda_{LAT}^2}} + \dots,
\end{equation}
which again reveals the vanishing of $g_0$ in the continuum limit:
\begin{equation}
g_0^2 \rightarrow 0 \quad \mbox{for} \ a \rightarrow 0.
\end{equation}
We can also observe that
\begin{equation}
am = C\, \exp \left( -\frac{1}{2\beta_0 g_0^2}\right) \cdot (\dots),
\label{amscaling}
\end{equation}
which shows the non-perturbative origin of the mass $m$.

These considerations, based on the perturbative $\beta$-function, motivate
the following hypothesis: the continuum limit of a gauge theory on a lattice
is to be taken at $g_0 \rightarrow 0$. Moreover, we expect that it involves
massive interacting glueballs and static quark confinement.

\begin{figure}[hbt]
\begin{center}
\epsfig{file=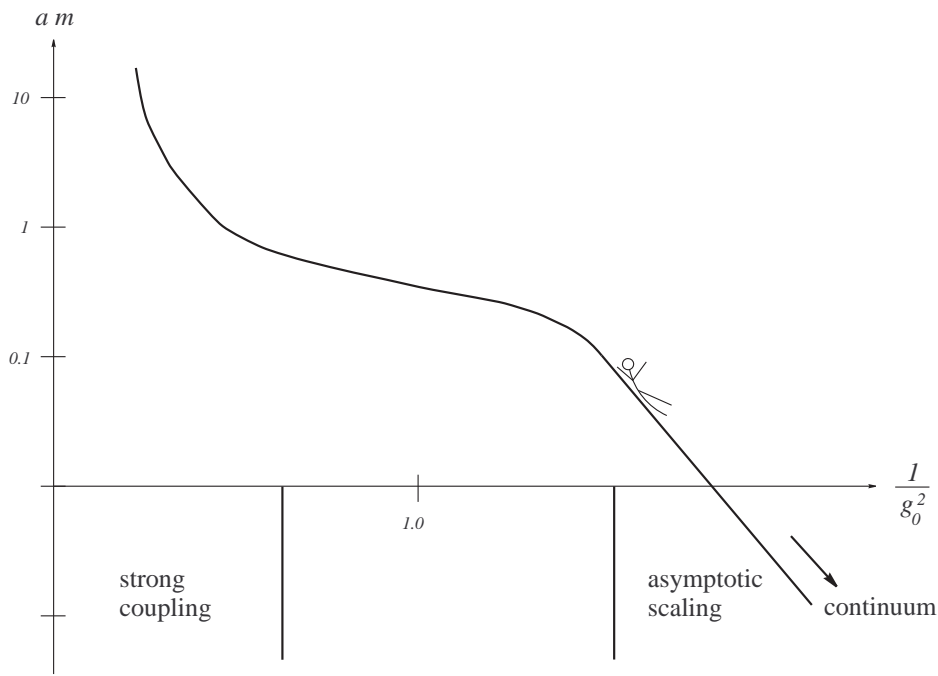,height=9.5cm}
\caption{Scaling scenario}
\label{scaling}
\end{center}
\end{figure}

The scenario for approaching the continuum limit then is as follows.
Calculating masses in lattice units, i.e.\ numbers $am$, and decreasing
$g_0$, we should reach a region where dimensionless quantities $am$
follow a behaviour as given by Eq.~(\ref{amscaling}). Plotting $am$
logarithmically versus $1/g_0^2$ as in Fig.~\ref{scaling}, an approximate
linear behaviour is expected, which is called {\em asymptotic scaling}.

For mass ratios it can be shown that the exponential dependence on $1/g_0^2$
cancels out and that near the continuum limit
\begin{equation}
\frac{m_1}{m_2} = \mbox{const.} \times (1+{\cal O}(a^2)).
\end{equation}
Such a behaviour, $m_1 / m_2 \approx$ const., is called {\em scaling}.

\subsection{Error Sources}

The results obtained by means of the Monte Carlo method differ from the
desired physical results by different sorts of errors. The most important
error sources are
\begin{itemize}
\item {\em statistical errors}: due to the finite number of
configurations in the Monte Carlo calculation,
$\sim 1 / n^{1/2}$,
\item {\em lattice effects}: due to finite lattice spacing $a$,
often $\sim a$ or $a^2$,
\item {\em volume effects}: due to finite lattice volume,
often $\sim 1 / L$, $1 / L^2$, or $e^{-mL}$,
\item {\em large quark masses}: $m_q$ mostly too big in
Monte Carlo calculations,
\item {\em quenched approximation}: $\det Q=1$, neglecting the fermion
dynamics.
\end{itemize}

\subsection{QCD Calculations}

In the lectures at Zuoz the methods, physical quantities and error sources
introduced have been illustrated by various numerical results for lattice
QCD, obtained by different groups in recent years. This includes results on
the
\begin{itemize}
\item glueball spectrum,
\item static quark potential,
\item flux tubes between static quarks,
\item hadron spectrum and decay constants,
\item QCD running coupling $\alpha_S$.
\end{itemize}

Since the corresponding figures can be found in the literature, we decided
not to reproduce them here and instead refer to the literature
\cite{results,reviews}.

Other topics of physical interest, which have been investigated in lattice
QCD, include hadronic and weak matrix elements, finite temperature QCD,
quark-gluon plasma etc..

\subsection{Higgs Models}

As already mentioned in the beginning, non-perturbative studies of gauge
theories are of interest in the electroweak sector as well. With Higgs
models (pure gauge theory with scalar Higgs field) the electroweak phase
transition has been studied. In Yukawa models (scalar fields and fermionic
fields) the dynamical mass generation due to the coupling of Higgs fields to
the fermions has been investigated in order to find bounds for the Higgs
mass. Studies with Higgs-Yukawa models, including all three types of fields,
have also been performed.

To conclude, a variety of lattice studies to extract parameters of the
standard model and to clarify implications of the physics beyond have been
made and are on their way.

\subsection*{Acknowledgements}

We would like to thank the organizers of the PSI Summer School for the
pleasant week in Zuoz, full of physics and fun in beautiful surroundings.
In particular we thank Dirk Graudenz for his efforts, which led the
school to success.


\end{document}